# A Novel Approach for Designing Online Testable Reversible Circuits

Md. Selim Al Mamun[1], Pronab Kumar Mondal[2], Uzzal Kumar Prodhan[3]

[1,2,3]Jatiya Kabi Kazi Nazrul Islam University, Trishal, Mymensingh, Bangladesh.

**Abstract:-** Reversible logic is gaining interest of many researchers due to its low power dissipating characteristic. In this paper we proposed a new approach for designing online testable reversible circuits. The resultant testable reversible circuit can detect any single bit error whiles it is operating. Appropriate theorems and lemmas are presented to clarify the proposed design. The experimental results show that our design approach is superior in terms of number of number of gates, garbage outputs and quantum cost.

## I. INTRODUCTION

Conventional logic dissipates a significant amount of energy because information bits are lost during logic operations. R. Launder[1] demonstrated that each bit of information loss dissipates kT*ln2 joules of heat energy, where k is Boltzmann's constant and T the absolute temperature at which computation is performed. Reversible logic is being considered as an alternative of traditional logic since reversible computing does not lose any information. According to Frank [2], reversible logic can recover a fraction of energy that can reach up to 100%. C.H. Bennett [3] proved that the energy loss problem can be avoided if the circuits are built using reversible gates. A reversible gate is a logical cell that has the same number of inputs and outputs, inputs and outputs have a one-to-one mapping [4]. Several reversible gates have been designed till date. The important basic reversible logic gates are Feynman gate [5] which is the only 2*2 reversible gate. There is also Toffoli gate [6], Frekdin gate [7], Peres gate [8], all of which can be used to realize important combinational functions and all are 3*3 reversible gates. Some popular reversible gates are shown in Fig.1. A good amount of research work has been carried out in the area of reversible logic and testing. However, there is not much work in literature on reversible logic with online testability. A circuit is called online testable if it is possible to test the circuit while the circuit is performing the operation. In this paper we proposed a simple efficient approach to design online testable reversible circuit that can detect any single bit error in the circuit. Our approach works with any circuit that is designed using reversible logic gates. The general idea is to convert each $n \times n$ reversible gate to a $(n+1) \times (n+1)$ testable block that is also reversible. The additional $(n+1)$th output bit is used to detect the error. This error bit of every testable block is carried to a proposed checker circuit, producing a final error bit. Thus examining the single error bit at the end, any single bit error can be detected. Comparisons of our design with the existing ones show the efficiency of our design in terms of number of gates, number of garbage outputs and quantum cost. The rest of the paper is organized as follows. Section 2 presents some related research on online testability of reversible circuits. Section 3 describes our proposed approach for construction of online testable reversible circuits. Section 4 presents our experimental results and compares our results with the existing ones in literature. Finally this paper is concluded with Section 5.

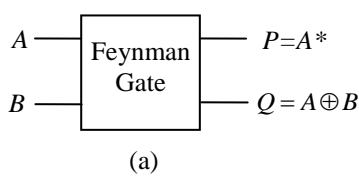
(a)

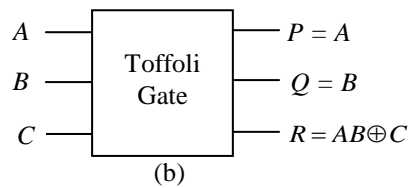
(b)

*A Novel Approach for Designing Online Testable Reversible Circuits*

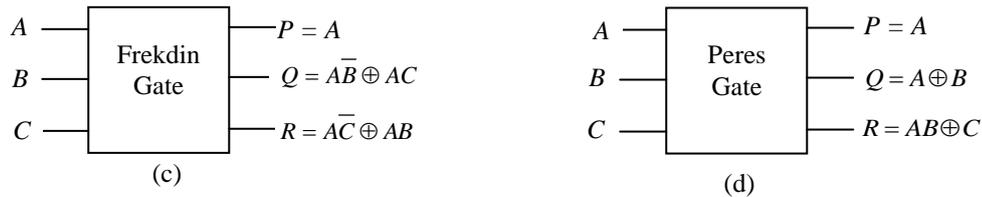

(c)                        (d)

**Fig.1.** (a) Feynman gate, (b) Toffoli gate, (c) Frekdin gate and (d) Peres gate

## II. RELATED WORKS

Researchers have worked on various aspects of reversible logic testing, such as fault modeling [9], test pattern generation [10] and many other areas. This section discusses a number of different approaches to generate online testable reversible circuits reported in the literature and explores the issues regarding their designs on error detection.

In [11][12], Vasudevan et al. proposed a design methodology to construct a reversible circuit with online testability. Two $4 \times 4$ reversible gates $R_1$ and $R_2$ and one $3 \times 3$ reversible gate $R$ were proposed. A combination of these $R_1$ and $R_2$ gates were used to construct testable logic blocks. The $R$ gate was used to construct a two pair rail checker, for detecting errors in any pair of testable blocks. Mahammad et al. [13, 14] proposed an extension of the previous approach. Each $n \times n$ reversible gate $G$ used in the circuit is converted into a $(n+1) \times (n+1)$ deduced reversible gate *DRG (G)*. This gate is cascaded to a special *DRG(X)* gate to construct Testable Reversible Cell (TRC) where X is an $n \times n$ reversible gate that has the same input and output vectors. To test every TRC a $(2n+1) \times (2n+1)$ bit error checker circuit called Testable Cell (TC) is formed. Thapliyal and Vinod [14] proposed an approach similar to the previous one [11][12]. A new $4 \times 4$ reversible Online Testable Gate (OTG) introduced in their work has a parity output at *q*. The $R_2$ gate is combined with the OTG to design a block with online testability feature. Two parity outputs *s* and *q* of this testable block are compared to check whether the block is faulty or not. In [11,12], the two-pair two-rail checker circuit was designed using six *R* gates, whereas in [15] the authors suggested a design using four $3 \times 3$ Toffoli gates and two $3 \times 3$ Fredkin gates.

## III. OUR APPROACH

In this section we presented our approach to convert a reversible circuit into an online testable reversible circuit that can detect any single bit error in the circuit. Appropriate lemmas and theorems are also presented to prove the correctness of our approach. Our proposed approach is described in 3 steps.

**Step 1: Convert each reversible gate in the circuit to Testable Reversible Gate (TRG).**

In this step an $n \times n$ reversible gate, $R$ is converted to $(n+1) \times (n+1)$ Testable Reversible Gate, TRG(R). Here, approach proposed by Mahammad et al. [13],[14] is used for the conversation. Let input vector $I_v = (I_1, I_2, ..., I_n)$ and output vector $O_v = (O_1, O_2, ..., O_n)$ of an $n \times n$ reversible gate R, then the Testable Reversible Gate, TRG(R) is defined as $I_v = (I_1, I_2, ..., I_n, 0)$ and $O_v = (O_1, O_2, ..., O_n, P_{out(TRG)})$ where $P_{out(TRG)} = O_1 \oplus O_2 \oplus ... \oplus O_n$. The conversion is shown in Fig 2.

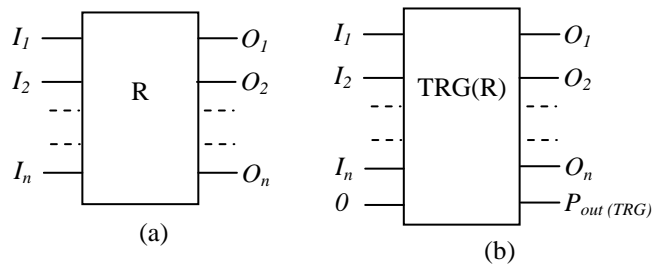

(a)                    (b)

**Fig.2.** (a) An $n \times n$ reversible gate R and (b) Testable Reversible Gate(TRG)

**Lemma 1:** TRG(R) is reversible.
**Proof:** If $I_v(I_1, I_2, ..., I_n)$ is mapped to $O_v(O_1, O_2, ..., O_n)$ in R then $I_v = (I_1, I_2, ..., I_n, 0)$ mapped to $O_v = (O_1, O_2, ..., O_n, O_1 \oplus O_2 \oplus ... \oplus O_n)$ in TRG(R). Hence TRG(R) is reversible.

**Step2: Construct Testable Block (TB) from Testable Reversible Gate (TRG).**

A Testable Block (TB) is constructed by cascading TRG(R) with a $(n+1) \times (n+1)$ TRG(S) gate as shown in Fig. 3 where S is an $n \times n$ gate that has the same input and output vectors.

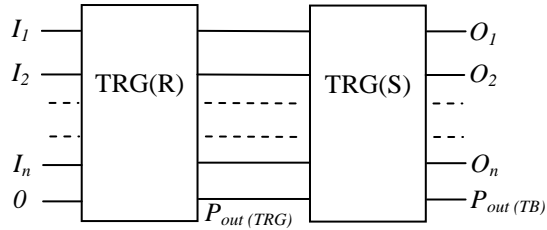

**Fig.3.** Testable Block (TB)

**Lemma 2:** TB(R) is reversible.
**Proof:** Testable Block (TB) is constructed from two reversible gates TRG(R) and TRG(S). Any circuit constructed from reversible circuit is also reversible; hence TB(R) is reversible.

**Lemma 3:** TB(R) is erroneous if $(n+1)^{th}$ output of TB(R) produces 1.
**Proof:** From the construction of TB(R), note that the $(n+1)^{th}$ output of TRG(R) is $P_{out(TRG)} = O_1 \oplus O_2 \oplus ... \oplus O_n$ which is the $(n+1)^{th}$ input of the TRG(S) gate. The $(n+1)^{th}$ output of TB(R) is $P_{out(TB)} = P_{out(TRG)} \oplus O_1 \oplus O_2 \oplus ... \oplus O_n = O_1 \oplus O_2 \oplus ... \oplus O_n \oplus O_1 \oplus O_2 \oplus ... \oplus O_n = 0$. So $(n+1)^{th}$ output is 0 if there is no error and 1 if there occurs an error.

Step 3: Construct error checker circuit for Testing Blocks (TB).

A modified 3×3 Frekdin gate proposed by Mamun et al in [16] is used as the building block for error checker circuit. The input vector, $I_v$ and output vector, $O_v$ for 3×3 modified Fredkin Gate (MFRG) is defined as follows: $I_v = (A, B, C)$ and $O_v = (P=A, Q = A\overline{B} \oplus AC, R = \overline{AC} \oplus AB)$. The quantum cost of MFRG gate is 4. Fig.4 shows the modified Frekdin gate MFRG.

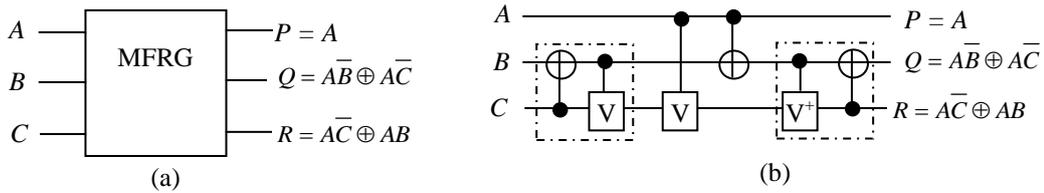

**Fig.4.** (a) Modified Frekdin Gate (MFRG) and (b) Equivalent quantum realization

All the $(n+1)^{th}$ output of TB(R) is carried out to the inputs of error checker circuit which is constructed cascading MFRG gates. Fig.5 shows the final online testable reversible circuits.

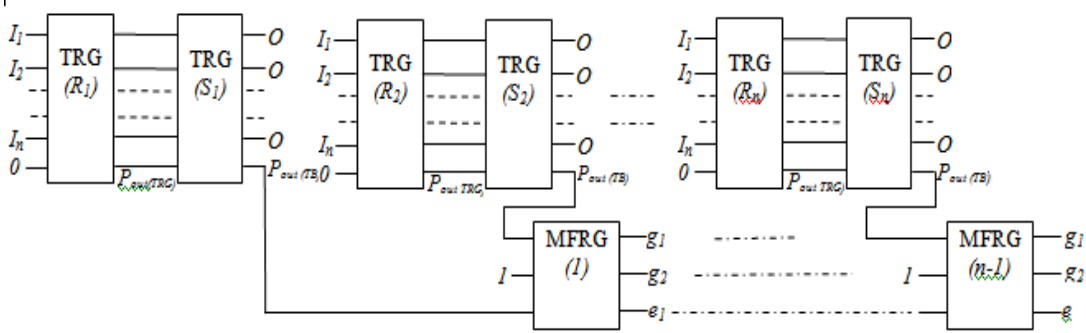

**Fig.5.** Block diagram of online testable reversible circuits.

**Theorem 1:** The circuit constructed in step 3 has the following two properties.
1) The circuit is reversible.
2) If there is a single bit error in any TB(R) then the output E = 1.

**Proof:**
1) From lemma 2 we get that all the testable blocks, TB(R)s are reversible. These reversible blocks are combined with a reversible circuit consisting of reversible MFRGs. Thus the resultant circuit is reversible.

2) From lemma 3 we get that if a testable block TB(R) is erroneous the $(n+1)^{th}$ output e =1. Let consider two testable blocks $TB_1(R)$ and $TB_{2(R)}$ with output $e_1$ and $e_2$ which are carried out to the $1^{st}$ and $3^{rd}$ inputs of the error checker circuit consisting of a single MFRG gate. The output $e = \overline{e1}.e2 \oplus e1 = e1+e2$. So if any of the $e_1$ and $e_2$ is 1 $e$ will be 1 which proves the theorem for two testable blocks.

For n testable blocks the output $e = e_1+e_2+.... +e_n$. If any of the n testable blocks produces error, $e$ output of the corresponding block will be *1*, hence $e = 1$ which proves the theorem for any number of testable blocks.

**Theorem 2:** n checker (can check n bits) circuit can be realized by at least *n-1* gates.
**Proof:** A single MFRG can be used for 2-checker circuit. We can systematically add additional *n-2* MFRG gates to produce n-checker circuit. Thus n-checker can be realized by at least *n-1* gates.

**Theorem 3:** n checker circuit can be realized by at least 2(n-1) garbage outputs.
**Proof:** A MFRG produces 2 garbage outputs. For n-checker circuit we need *n-1* MFRG gates. So in total *2(n-1)* garbage outputs are produced in the construction of n checker circuit.

**Theorem 4:** A reversible circuit of n reversible gates can be realized by at least *3n-1* gates.
**Proof:** For every reversible gate in the circuit a testable block TB(R) is constructed which consists of two gates. For n testable block we need an n-checker circuit consisting of *n-1* gates. So total number of gates = *2n+n-1 = 3n-1*.

**Theorem 5**: n testable block-checker circuit can be realized by least *2(n-1)* garbage outputs.
**Proof:** For every n×n reversible gate a testable block consisting of two *(n+1) × (n+1)* TRG(R) and TRG(S) gates are constructed. All the *n+1* outputs of the TRG(R) are used as the *n+1* input of TRG(S) gates and all the outputs of TRG(S) are primary outputs. So no garbage output is produced for testable block. From theorem 3 we know *2(n-1)* garbage outputs are produced for n-checker circuit. So in total *2(n-1)* garbage outputs are produced.

## IV.     EXPERIMENTAL RESULTS AND DISCUSSIONS

In this section we discussed the outcome and contribution of our paper by comparing our results with the existing ones that are available in literature.

### A.    Evaluation of Proposed Checker Circuit:

We proposed a checker circuit that that uses only modified Frekdin gate for two testable blocks. For *n* testable blocks we need only *n-1* modified Frekdin gates. The comparisons of our proposed design of checker circuit with the existing ones in literature are summarized in table I.

**Table I:** Comparisons of different checker circuits

| Design of n-checker circuit for n testable blocks | Cost Comparisons | | |
|---|---|---|---|
| | No. of gates | Garbage outputs | Quantum cost |
| **Proposed** | n-1 | 2(n-1) | 4(n-1) |
| Existing[12] | 6n | 8n | Quantum realization of R gate is not available |
| Existing [15] | 6n | 8n | 30n |
| Existing[17] | 2n | 3n | 6n |

### B.    Evaluation of proposed approach

Here we presented our proposed approach for designing online testable reversible circuit. The comparisons of our proposed approach with the existing ones in literature are summarized in table II.

**Table II:** Comparisons of different design approaches of online testable circuits

| Design of circuits of n Testable Blocks | Cost Comparisons | |
|---|---|---|
| | Number of Gates | Number of Garbage Outputs |
| **Proposed** | 3n-1 | 2(n-1) |
| Existing[12] | 8n | 10n |
| Existing[15] | 8n | 10n |
| Existing[17] | 4n | 5n |

In this paper we not only highlighted on the available designs but also implemented some popular benchmark functions. A number of benchmark functions collected from [18] are also implemented using approach. In table III we presented the comparisons in terms of number of gates and garbage outputs. The implementation results of [12] and [17] are taken from [17].

**Table III:** Comparisons of proposed approach with existing ones for benchmark functions

| Benchmark with online testability | Number of Gates | | | Number of Garbage Outputs | | |
|---|---|---|---|---|---|---|
| | Existing[12] | Existing[17] | **Proposed** | Existing[12] | Existing[17] | **Proposed** |
| Hwb4 | 178 | 92 | 32 | 208 | 101 | 20 |
| Mod5adder | 490 | 248 | 44 | 587 | 285 | 28 |
| Xor5 | 26 | 16 | 11 | 32 | 20 | 10 |
| Ham7 | 386 | 196 | 74 | 451 | 214 | 48 |
| 4mod5 | 66 | 36 | 14 | 79 | 42 | 12 |
| Rd32 | 82 | 44 | 11 | 98 | 51 | 8 |
| 5mod5 | 306 | 156 | 23 | 368 | 181 | 19 |
| 4 49 | 242 | 124 | 35 | 286 | 139 | 22 |
| Hwb5 | 578 | 292 | 71 | 686 | 329 | 46 |
| Rd53 | 306 | 156 | 35 | 367 | 180 | 26 |
| Ham3 | 66 | 36 | 11 | 76 | 39 | 6 |
| 3 17 | 106 | 56 | 17 | 125 | 63 | 10 |

From the experimental results it is found that performance of our experimental results is superior than that of the existing design in terms of number of gates and number of garbage outputs.

## V. CONCLUSION

Testing is required to ensure the quality and reliability of a circuit. Testing reversible circuit is very challenge as the levels of logic are significantly higher than the standard logic. In this paper we proposed a new approach for the realization of online reversible testable circuit that is efficient in terms of number of gates, delay and garbage outputs . While describing our approach we presented appropriate theorems and lemmas to prove the correctness of our approach. Comparative studies and experimental results of benchmark functions reflect the superiority of our design.

**ACKNOWLEDGMENT**

The authors would like to thank the anonymous referees for their constructive feedback, which helped significantly improving technical quality of this paper
.